\documentclass[11pt, a4paper]{article}

\usepackage{graphicx}
\usepackage{epsfig}

\newcommand\lesssim{\mathrel{\rlap{\lower4pt\hbox{\hskip1pt$\sim$}}
    \raise1pt\hbox{$<$}}}
\newcommand\gtrsim{\mathrel{\rlap{\lower4pt\hbox{\hskip1pt$\sim$}}
    \raise1pt\hbox{$>$}}}
\begin{document}

\thispagestyle{empty}

\title{Sachs-Wolfe effect in some anisotropic models}

\author{\footnotesize Paulo Aguiar \footnote{paguiar@cosmo.fis.fc.ul.pt} \hspace*{1mm} and
        \footnotesize Paulo Crawford \footnote{crawford@cosmo.fis.fc.ul.pt} \\
        \small Centro de Astronomia e Astrof\'\i sica da Universidade de Lisboa \\
        \small Campo Grande, Ed. C8; 1749-016 Lisboa, Portugal}

\date{}

\maketitle

\begin{abstract}
In this work it is shown for some spatially homogeneous but
aniso\-tropic models how the inhomogeneities in the distribution of
matter on the surface of the last scattering produce anisotropies
in large angular scales (larger than $\vartheta \gtrsim 2^\circ$)
which do not differ from the ones produced in
Friedmann-Lema\^{\i}tre-Robertson-Walker (FLRW) geometries. That
is, for these anisotropic models, the imprint left on the cosmic
microwave background radiation (CMBR) by the primordial density
fluctuations, in the form of a fractional variation of the
temperature of this radiation, is governed by the same expression
as the one given for FLRW models. More precisely, under adiabatic
initial conditions, the classical Sachs-Wolfe effect is recovered,
provided the anisotropy of the overall expansion is small. This
conclusion is in agreement with previous work on the same
anisotropic models where we found that they may go through an
`isotropization' process up to the point that the observations are
unable to distinguish them from the standard FLRW model, if the
Hubble parameters along the orthogonal directions are assumed to
be approximately equal at the present epoch. Here we assumed upper
bounds on the present values of anisotropic parameters imposed by
COBE observations.
\end{abstract}

\section{Introduction}

The task of proving the homogeneity and isotropy of the Universe
at large scales is not a simple one. It is generally accepted that
the Universe is spatially homogeneous as a result of the so called
Copernican principle, that is, the assumption that we live in a
typical place. Then, a fundamental result follows from this
principle that if the CMBR temperature were exactly isotropic
about our position, it should be exactly isotropic about every
point in spacetime, and the universe would have to be exactly an
FLRW model. This was proved by Ehlers, Geren and Sachs \cite{EGS},
and known as the EGS theorem.

As a matter of fact, observations indicate that the temperature of
the CMBR is isotropic to a remarkable degree. This observation,
put together with the Copernican principle, leads to the
widespread believe that the Universe can be accurately described
by a spatially homogeneous and isotropic model on sufficiently
large scales, drastically reducing the space of solutions of
Einstein equations, and the number of possible cosmological
models.

However, the EGS theorem is not directly applicable to the real
Universe since the CMBR temperature is not exactly isotropic. This
fact might explain why, despite the high level of isotropy of CMBR
temperature, some authors have worked on spatially homogeneous but
anisotropic models to analyze whether they might agree with
present observations, in particular with those models which could
be considered to be close to FLRW. Recall that one defines to be
`close' to a FLRW model when both parameters
\begin{equation}
{\cal W}^2= \frac{E_{ab}E^{ab}+H_{ab}H^{ab}}{6H^4}, \label{Weyl}
\end{equation}
\begin{equation}
\Sigma^2=\frac{\sigma_{ab}\sigma^{ab}}{6H^2} \label{Si}
\end{equation}
are almost zero, although the former usually receives more
attention than the latter in the astrophysical literature, as was
stressed in \cite{Nilsson}. Here, $\sigma_{ab}$ is the shear
tensor, $H$ is the mean Hubble parameter, $E_{ab}$ and $H_{ab}$
are the electric and magnetic part of the Weyl tensor,
respectively, and we refer to
$\Sigma$ as the shear parameter and to ${\cal W}$ as the Weyl
parameter. Note that for non-tilted spatially homogeneous perfect
fluid models, a zero shear tensor implies that the Weyl curvature
tensor is zero, and thus characterizes the FLRW models, that is,
$\Sigma=0 \Rightarrow {\cal W}=0$. However, restricting $\Sigma$
to be small does not guarantee that ${\cal W}$ is small since the
Weyl curvature tensor is related to time derivatives of the shear
tensor and these need not be small compared to $H^2$. It is thus
of considerable interest to note that the EGS theorem might be
extended as was done in \cite{Stoeger}, given some reasonable
assumptions, by replacing the word ``exactly" by ``almost", thus
obtaining an ``almost" EGS theorem. In other words, they showed,
given certain assumptions, that if the CMBR temperature is
measured to be almost isotropic in a spacetime region of an
expanding universe, then the universe is close to an FLRW model in
that region. In terms of the previously introduced anisotropy
parameters, the condition for the universe to be close to an FLRW
model becomes
\begin{equation}
\Sigma \ll 1, \quad {\cal W}\ll 1. \label{anisotropy}
\end{equation}

It should be stressed, however, that in \cite{Lim} it is shown
that there are spatially homogeneous cosmological models such that
the CMBR temperature is measured to be isotropic at a given time
$t_0$ by all fundamental observers, even though the overall
expansion of the universe is highly anisotropic at $t_0$, here
assumed to be present time.

On the other hand, if the classical tests of cosmology are applied
to a simple Kantowski-Sachs metric and the results compared with
those obtained for the standard model, the observations will not
be able to distinguish between these models if the Hubble
parameters along the orthogonal directions are assumed to be
approximately equal \cite{Henriques} at $t_0$, that is, if $\Sigma
\approx 0$. Following along the same lines, we made a qualitative
study \cite{Aguiar} of three axially symmetric metrics
(Kantowski-Sachs, Bianchi type-I and Bianchi type-III), with a
cosmological constant and dust, to analyze which were physically
permitted, when we assume them to be bound by a high degree of
isotropy, from the point of view of overall expansion. More
specifically, although our geometries described anisotropic
cosmological fields they could be considered to be almost FLRW as
far as its overall expansion, since the shear tensor was close to
zero, at least since the time of last scattering. We found that
these models undergo `isotropization' up to the point that the
observations will not be able to distinguish between them and the
standard model, except for the Kantowski-Sachs model
$(\Omega_{k_{0}}<0)$ and for the Bianchi type-III model
$(\Omega_{k_{0}}>0)$ with $\Omega_{\Lambda_{0}}$ smaller than some
critical value $\Omega_{\Lambda_{M}}$ \cite{Aguiar} (see the
definitions of $\Omega_{k_{0}}, \Omega_{\Lambda_{0}},
\Omega_{{M_0}}$ in \cite{Aguiar2}). From this analysis we concluded
that these models are good candidates for the description of the
observed Universe, provided that the Hubble parameters are
approximately equal at the last scattering (see computations
below). In other words, the low values of the first parameter,
$\Sigma$, is sufficient to assure a FLRW-like behavior, and the
statement that the expansion is highly isotropic means that
$\Sigma \ll 1$.

Historically, the detection of the CMBR has led to constrains in
theoretical models in the field of Cosmology, and favored the Big
Bang solutions. Indeed, it was the observed level of isotropy of
the CMBR temperature, first detected by Penzias \& Wilson
\cite{Penzias}, which has been considered to provide the best
evidence for the large-scale isotropy of the Universe, and still
is the strongest argument in favor of an isotropically expanding
Universe. Later, more precise experiments proved that this
radiation has temperature fluctuations, or anisotropies. These
small anisotropies are thought to give rise to the observed
galaxies, and large-scale structures in the Universe.

In 1992, the COBE (COsmic Background Explorer) satellite
\cite{Smoot,Coles} observed the CMBR with unprecedented precision
and revealed for the first time that the level of the CMBR
temperature fluctuations on large scales is as small as ${{\Delta
T} \over T}\simeq 10^{-5}$ \cite{Mather,PartridgeCOBE}. After COBE
many other ground and balloon born experiments \cite{Partridge},
with higher angular resolution, confirmed this result and allowed
us to probe the level of the anisotropies on a large range of
scales.

On large angular scales, the CMBR anisotropies ($\frac{\Delta
T}{T}$), are dominated by the Sachs-Wolfe effect. This phenomenon,
already deduced theoretically by Sachs \& Wolfe \cite{Sachs}, was
used to compute the first-order perturbations in a FLRW universe
with a flat 3-space filled either with dust or radiation. This is
just one of the various possible sources of anisotropy, which
occurs when there are inhomogeneities in the distribution of
matter on the surface of the last scattering, that may produce
anisotropies by the redshift or blueshift of photons. In this
paper we compute the Sachs-Wolfe effect \cite{Sachs} for some
anisotropic but homogeneous models (Kantowski-Sachs and Bianchi
type-III model which is also locally rotationally symmetric (LRS))
and find that under the assumption $H_{a_0} \simeq H_{b_0}$ (we
considered a small anisotropy imposed by COBE observations: see
\cite{Aguiar2} for details) these models allow us to recover the
classic Sachs-Wolfe effect obtained for FLRW universes. This is an
interesting result which tells us that CMBR observations on large
angular scales will not be able to distinguish these anisotropic
models from FLRW ones.

\section{The method}

As Collins and Hawking \cite{Collins} pointed out, the number of
cosmological solutions which demonstrate exact isotropy well after
the Big Bang origin of the Universe is a small fraction of the set
of allowable solutions to the Einstein equations. It is therefore
prudent to take seriously the possibility that the Universe is
expanding anisotropically and to investigate what effect
anisotropic expansion will have on the angular distribution of
background radiation \cite{Partridge}. In this work we show that,
for large angular scales ($\vartheta \gtrsim 2^\circ$), there
exist homogeneous but anisotropic models, where the photons
travelling to an observer from the last scattering surface
encounter metric perturbations which cause them to change
frequency, just like in the case of FLRW models.

The metrics we consider are Kantowski-Sachs and LRS Bianchi
type-III model, given by
\begin{equation}
d \tilde
s^2=-dt^2+a^2(t)dr^2+b^2(t)(d\theta^2+f^2(\theta)d\phi^2),
\label{metr1}
\end{equation}
where
$$
f(\theta)= \left\{ \begin{array}{ll}
\sin \theta & \mbox{for Kantowski-Sachs} \\
\sinh \theta & \mbox{for Bianchi type-III}
\end{array} \right.
$$
We evaluate the Sachs-Wolfe effect \cite{Sachs,White}, assuming
small perturbations in the previous metrics, and then integrating
the geodesic equations for the CMBR photons along their paths,
from the Last Scattering Surface (LSS) to the observer. In this
work we account for the ``kinematics effects" undergone by the
free propagating radiation from the last scattering, in a
perturbed universe, and for the ``intrinsic effects" originated by
the set of physical and microphysical processes related to the
density perturbations in the LSS. For simplicity, it is common to
perform a conformal transformation \footnote{Owing to this
transformation, $\eta$ is usually called the conformal time, and
it is related to cosmological proper time by $dt^2=d \eta^2
a^2(\eta)$.} of the previous metrics, and to work with the
following metric forms
\begin{equation}
d \bar {\tilde s}^2= -d\eta^2+ dr^2+\frac{b^2(\eta)} {a^2(\eta)}
(d\theta^2+f^2(\theta)d\phi^2) \label{metr2}
\end{equation}
such that $d \bar {\tilde s}^2 $: $ d \tilde s^2 =a^2(\eta) d \bar
{\tilde s}^2$, since the null geodesics are preserved by this
transformation. Afterwards, the results are transported to $d
\tilde s^2$ metric. The metric $ d \bar {\tilde s}^2$ is perturbed
in the following way
\begin{eqnarray}
d \bar s^2 &=& -(1+ h_{00})d\eta^2+ (1+h_{11})dr^2+
\frac{b^2(\eta)}{a^2(\eta)} \left[(1+h_{22})d\theta^2 \right. \nonumber \\
& & \left. +(1+h_{33})f^2(\theta)d\phi^2 \right]
-(h_{01}+h_{10})d\eta dr
-\frac{b(\eta)}{a(\eta)}(h_{02}+h_{20})d\eta d\theta \nonumber \\
& &-\frac{b(\eta)}{a(\eta)}f(\theta)(h_{03}+h_{30})d\eta d\phi
+\frac{b(\eta)}{a(\eta)}(h_{12}+h_{21})dr d\theta \nonumber \\
& & +\frac{b(\eta)}{a(\eta)}f(\theta)(h_{13}+h_{31})dr d\phi
+\frac{b^2(\eta)}{a^2(\eta)}f(\theta)(h_{23}+h_{32})d\theta d\phi,
\end{eqnarray}
where $h_{ab}$ are functions of time and position and such that
$h_{ab}\ll 1$.

Considering the geodesics equation for the photons
\begin{equation}
{{d \bar U_a} \over {dw}}={1 \over 2} \bar g_{bc,a}\bar U^b \bar U^c,
\end{equation}
where $\bar U^a$ represents the photon 4-vector velocity components and
$w$ the affine parameter associated to its trajectory, we may
calculate this 4-velocity integrating the previous equation
\begin{equation}
\bar U_a={1 \over 2}\int {\bar g_{bc,a}}\bar U^b \bar U^cdw+{}_{(0)}\bar U_a.
\end{equation}
The term ${}_{(0)} \bar U_a$ represents the non perturbed photon
4-velocity components in the covariant form.

A material observer, moving with some 3-velocity $\vec {\bar V}$, in a
perturbed universe (metric $\bar g_{ab}$), has a 4-velocity given by
\begin{equation}
\bar V^a \simeq \left(1-{1 \over 2}h_{00},\bar V^1,\bar V^2,\bar V^3 \right).
\end{equation}
This observer measures a photon energy $\bar E_{\gamma}$
proportional to $\bar U_a \bar V^a$,
\begin{equation}
\bar E_\gamma \propto \left( {{1 \over 2}\int {\bar
g_{bc,a}\bar U^b \bar U^cdw+{}_{(0)}\bar U_a}} \right)
\left( {1-{1 \over 2}h_{00},\bar V^i} \right).
\end{equation}
Decomposing the 4-velocity in a Taylor expansion and conserving
only the first order term $\bar U^a \simeq {}_{(0)}\bar U^a +{}_{(1)}\bar U^a$
(it can be shown that the 4-velocity, in covariant and contravariant components, is given by
${}_{(0)}\bar U^a=(1, \sqrt\alpha,a^2/b^2\sqrt\beta,0)$ and
${}_{(0)}\bar U_a=(-1, \sqrt\alpha, \sqrt\beta,0)$)
and neglecting terms like $h_{00}{}_{(1)}\bar U^a$ (because they are
second order terms), the expression becomes
\begin{eqnarray}
\bar E_\gamma & \propto & {1 \over 2}\int \bar g_{bc,0}\bar U^b \bar U^cdw
-{1 \over 4}h_{00} \int \bar g_{bc,0}{}_{(0)}\bar U^b{}_{(0)}\bar U^cdw
+{}_{(0)}\bar U_0 \nonumber
\\ & &
- {1 \over 2}{}_{(0)}\bar U_0 h_{00} + \vec {\bar U} \cdot \vec {\bar V} +{1 \over 2}
\int \bar g_{bc,i}{}_{(0)}\bar U^b{}_{(0)}\bar U^cdw \bar V^i.
\end{eqnarray}
The last term of the right hand side vanishes, because for $i=r$
and $i=\phi \Rightarrow \bar g_{bc,i}=0$ and also because
$\bar U^{\phi}=0$. The term
\begin{eqnarray}
& & {1 \over 4}h_{00} \int \bar g_{bc,0}{}_{(0)}\bar U^b{}_{(0)}\bar U^cdw =
{1 \over 4}h_{00} \int \bar {\tilde
g}_{bc,0}{}_{(0)}\bar U^b{}_{(0)}\bar U^cdw \nonumber
\\ & &
={1 \over 4}h_{00} \int \bar {\tilde g}_{\theta \theta, \eta}
\left({}_{(0)}\bar U^{\theta }\right)^2dw = {1 \over 2}h_{00}\int {b^2
\over a^2}\left({\dot b \over b}- {\dot a \over
a}\right)\left({}_{(0)}\bar U^{\theta }\right)^2dw, \nonumber
\end{eqnarray}
may be numerically computed. Here the dot represents a derivative
with respect to conformal time: $(~\dot{}~) \equiv \frac{d}{d
\eta}$. If we choose accurately the values of density parameters
$\Omega_M$ and $\Omega_{\Lambda}$ (see Aguiar \& Crawford
\cite{Aguiar}) such that $\Omega_M +\Omega_{\Lambda} \simeq 1$
this integral may be neglected, because it is a first order term
times another first order quantity (see \cite{Aguiar2}). The first
term in the right hand side may be decomposed by the following way
\begin{eqnarray}
& &{1 \over 2} \int \bar  g_{bc,0}\bar U^b\bar U^cdw \nonumber
\\ & &
={1 \over 2} \int \bar  g_{bc,0}{}_{(0)}\bar U^b{}_{(0)}\bar U^cdw + {1
\over 2} \int \bar  g_{bc,0}{}_{(0)}\bar U^b{}_{(1)}\bar U^cdw + {1 \over 2}
\int \bar  g_{bc,0}{}_{(1)}\bar U^b{}_{(0)}\bar U^cdw \nonumber
\\ & &
={1 \over 2} \int \bar  g_{bc,0}{}_{(0)}\bar U^b{}_{(0)}\bar U^cdw + \int
\bar {\tilde
g}_{\theta\theta,\eta}{}_{(0)}\bar U^{\theta}{}_{(1)}\bar U^{\theta}dw.
\nonumber
\end{eqnarray}
But
$$
\int \bar {\tilde
g}_{\theta\theta,\eta}{}_{(0)}\bar U^{\theta}{}_{(1)}\bar U^{\theta}dw
=2\int {b^2 \over a^2}\left({\dot b \over b}-{\dot a \over
a}\right) {}_{(0)}\bar U^{\theta}{}_{(1)}\bar U^{\theta}dw,
$$
may also be neglected for the reason pointed previously. Thus,
\begin{equation}
\bar E_\gamma \propto {1 \over 2}\int \bar
g_{bc,0}{}_{(0)}\bar U^b{}_{(0)}\bar U^cdw +{}_{(0)}\bar U_0 -{1 \over
2}{}_{(0)}\bar U_0h_{00}+\vec {\bar U} \cdot \vec {\bar V}.
\end{equation}
Calculating the ratio of energies in the emission instant ($e$)
and in the reception instant ($r$) we obtain
\begin{equation}
{\bar E_{\gamma_{e}} \over \bar E_{\gamma_{r}}}= {\left.( \bar
U_a \bar V^a)\right|_e \over \left.( \bar U_a \bar V^a)\right|_r}= {-1+{1\over
2}\left. h_{00}\right|_e + \left.(\vec {\bar U} \cdot \vec
{\bar V})\right|_e \over -1+{1\over 2}\left. h_{00}\right|_r +
\left.(\vec {\bar U} \cdot \vec {\bar V})\right|_r +{1 \over
2}\int\limits_e^r {\bar g_{bc,0}{}_{(0)}\bar U^b{}_{(0)}\bar
U^cdw}},
\end{equation}
(the symbol $\left. X \right|_A$ means that $X$ is being evaluated
at point $A$). Considering the approximation $(1+X)^{-1}\simeq
1-X$ we have
\begin{equation}
{\bar E_{\gamma_{e}} \over \bar E_{\gamma_{r}}}= 1+ {1 \over
2}\left[ h_{00}\right]_e^r+\left[ \vec {\bar U} \cdot \vec {\bar V} \right]_e^r
+{1 \over 2}\int\limits_e^r {\bar
g_{bc,0}{}_{(0)}\bar U^b{}_{(0)}\bar U^cdw},
\end{equation}
where $\left[ X \right]_A^B \equiv X(B)-X(A)$. The result obtained
in $d \bar s^2$ may be transported to $ds^2$ using the relation
$E(t)={1 \over a(t)} \bar E(w)$, as can easily be obtained.
The photon redshift between emission at the point $e$ in the LSS
and reception at the point $r$ is given by the ratio of the
measured energies at emission and reception,
\begin{equation}
z+1={\lambda_r \over \lambda_e}={E(t_e) \over E(t_r)}={a_r \over
a_e} {\bar E(w_e) \over \bar E(w_r)}. \label{shift}
\end{equation}
On the other hand, the redshift is also given by the ratio of the
black body associated temperatures at the emission and reception
times,
\begin{equation}
{T_e \over T_r}=z+1.
\end{equation}
Taking these last two equations, and considering again a linear
approximation one obtains
\begin{eqnarray}
T_r  & \simeq  & {a_e \over a_r} T_e \left[1- {1 \over 2}\left[
h_{00}\right]_e^r -\left[ \vec {\bar U} \cdot \vec {\bar V} \right]_e^r -{1
\over 2} \int\limits_e^r {\bar g_{bc,0}{}_{(0)}\bar U^b{}_{(0)}\bar U^cdw}
\right]
\nonumber \\
& = & {a_e \over a_r} T_e \left(1+ {\delta T_{\mbox{\tiny
journey}} \over T_{\mbox{\tiny journey}}} \right),
\end{eqnarray}
so to evaluate the observed anisotropy we must take into account
the journey and emission perturbations (see \cite{Liddle} p.357),
and write it as
\begin{equation}
{\delta T_r \over T_r}= {\delta T_e \over T_e}+ {\delta T_{\mbox{\tiny journey}}
\over T_{\mbox{\tiny journey}}}
\end{equation}
or equivalenty,
\begin{equation}
{\delta T_r \over T_r}={\delta T_e \over T_e} -{1 \over 2}\left[ h_{00}\right]_e^r
-\left[ \vec {\bar U} \cdot \vec {\bar V} \right]_e^r -{1 \over 2}
\int\limits_e^r {\bar g_{bc,0}{}_{(0)}\bar U^b{}_{(0)}\bar U^cdw}.
\label{deltaT}
\end{equation}

Now let us define the perturbations $h_{ab}$. These fluctuations
are {\em gauge} dependent. This means that one must make a gauge
choice which involves fixing the constant-time hypersurfaces (and
the spatial grid on these surfaces) where these fluctuations are
defined \cite{Mukhanov,Hu}. As it is widely assumed, we will
choose a conformal {\em Newtonian gauge} to allow a more intuitive
understanding. Considering only scalar perturbations, the non zero
quantities are (see \cite{Mukhanov} Equations (2.9) and (3.21),
$B=E=0$ for the longitudinal gauge)
\begin{equation}
h_{00}=2 \Psi; ~~ h_{11}= h_{22}=h_{33}=2 \Phi.
\end{equation}
As it was previously stated, these scalar quantities are functions
of time and position, $\Psi=\Psi(\eta,x^i)$, $\Phi=\Phi(\eta,x^i)$
and they may be seen as a Newtonian potential and a spatial
curvature perturbation potential, respectively \cite{Mukhanov,Hu}.

For our models, we have perfect fluid, so the anisotropic pressure is null.
This implies that $\Psi= - \Phi$ (see \cite{Hu} for details).
Writing (\ref{deltaT}) in the Newtonian gauge we obtain,
\begin{eqnarray}
{\delta T_r \over T_r} & = & {\delta T_e \over T_e} -\left[ \Psi \right]_e^r - \left[
\vec {\bar U} \cdot \vec {\bar V} \right]_e^r -{1 \over 2}\int\limits_e^r
\left\{\left[-h_{00,0}({}_{(0)}\bar U^{\eta})^2+ h_{11,0}({}_{(0)}\bar U^r)^2 \right. \right. \nonumber \\
& & \left. \left. +\frac{b^2}{a^2}h_{22,0}({}_{(0)}\bar U^{\theta})^2 \right]dw\right\}
 -\int\limits_e^r (1+h_{22}){{b^2 \over a^2} \left({\dot b \over
b}-{\dot a \over a}\right) ({}_{(0)}\bar U^{\theta})^2}dw.
\end{eqnarray}
The last term of the right hand side may be neglected, as it is
shown in \cite{Aguiar2}. So, we get
\begin{eqnarray}
{\delta T_r \over T_r} &=& {\delta T_e \over T_e} -\left[ \Psi \right]_e^r - \left[ \vec {\bar U}
\cdot \vec {\bar V} \right]_e^r \nonumber \\
&& -\int\limits_e^r {\left\{-{\partial \Psi
\over \partial \eta} ({}_{(0)}\bar U^{\eta})^2+{\partial \Phi \over
\partial \eta} \left[({}_{(0)}\bar U^r)^2
+\frac{b^2}{a^2}({}_{(0)}\bar U^{\theta})^2 \right] \right\}}dw,
\label{brack}
\end{eqnarray}
or,
\begin{equation}
{\delta T_r \over T_r} ={\delta T_e \over T_e} -\left[ \Psi \right]_e^r - \left[ \vec {\bar U}
\cdot \vec {\bar V} \right]_e^r +2\int\limits_e^r {{\partial \Psi \over
\partial \eta}}dw, \label{deltaT1}
\end{equation}
as can easily be shown. Now, we should spell out the
physical interpretation of each one of the three factors of the
right hand side of previous equation. When matter and radiation
are decoupled, free CMBR photons, climbing the gravitational
potential generated by density perturbations, undergo a
gravitational redshift, with corresponding loss of energy. The
photon energy variation in this process is given by the term
$\left[ \Psi \right]_e^r \equiv \Psi(r)-\Psi(e)$. The second term,
$\left[ \vec {\bar U} \cdot \vec {\bar V} \right]_e^r \equiv \vec
{\bar U} \cdot \left(\vec {\bar V}(r) - \vec {\bar V}(e)\right)$,
corresponds to the Doppler effect induced by the relative motion
of the observer in the emission and reception events. The Doppler
term has an observational meaning of a dipolar anisotropy ($\vec
{\bar U} \cdot \vec {\bar V}(r)$) on CMBR temperature and is
usually removed from the equation and dealt with separately. The
other term ($\vec {\bar U} \cdot \vec {\bar V}(e)$) can be dropped
for large angular scales (for details see \cite{Liddle} p.124 and
p.84). The last term tells us that the perturbing potential may
vary between the emission and reception instants.

For flat FLRW models without cosmological constant, $\Psi$ is time
constant \cite{White,Mukhanov}, so the last term in the right hand
side of Equation (\ref{deltaT1}) vanishes. In our case the
cosmological constant is not vanishing, and indeed $\Lambda$ plays
an important role in our analysis. Taking into account recent
observations \cite{Perl,Riess} which suggest $\Omega_0 \sim 0.3$
and $\Omega_{\Lambda_{0}} \sim 0.7$, we consider values such that
$\Omega_0+\Omega_{\Lambda_{0}} \simeq 1$ (see \cite{Aguiar2}).

Now, we may explicitly compute the intrinsic temperature
fluctuations $\delta T_e /T_e$, originated by the set of physical
and microphysical processes, associated to density perturbations
in LSS. Despite its youth, the Universe is already highly
isotropic (shear $\Sigma \ll 1$). Then, for simplicity, we assume
in this section that the Universe might be characterized by a flat
FLRW model. Because the density fluctuations are very small, we
may treat them in the context of linear theory of perturbations
using Stefan-Boltzmann law
\begin{equation}
\rho_{\gamma}=\sigma_{SB} T^4,
\end{equation}
where $\sigma_{SB}$ is a constant and $\rho_{\gamma}$ is the
radiation density. Differentiating this equation one easily
obtains
\begin{equation}
{\delta T_e \over T_e}={1 \over 4}{\delta \rho_{\gamma} \over
\rho_{\gamma}}.
\end{equation}
At this time, when the Universe is very young, the total energy
density is not only due to radiation. The baryonic matter plays an
identical role, and so the matter density $\rho_m$ is related with
$\rho_{\gamma}$ by
\begin{equation}
{\delta \rho_m \over \rho_m} -{3 \over 4}{\delta \rho_{\gamma}
\over \rho_{\gamma}}=0,
\end{equation}
if the perturbation mode is adiabatic and on scales larger than
the horizon at this time \cite{Kolb}. Then, for perturbations on
scales greater than the horizon we may write
\begin{equation}
{\delta T_e \over T_e}= {1 \over 3}{\delta \rho_m \over \rho_m}.
\end{equation}
From the last expression we see that, for adiabatic perturbations,
the over-density (under-density) regions are intrinsically hotter
(colder) than the LSS mean temperature. According with
\cite{White,Mukhanov}, $\delta \rho_m / \rho_m=-2 \Psi + {\mathcal
O} [(k/H)^2]$, where $k$ is the momentum associated to
perturbation scale and $H$ the Hubble
parameter\footnote{Nevertheless our models have two Hubble
parameters $H_a$ and $H_b$, they have practically the same value
$H_a \simeq H_b \simeq H$ due to our parameters choose.}. The
larger is the scale, the smaller is $k$, so, for perturbations
greater than the horizon $k \ll H$, the over-density locals
coincide with the potential well, because,
\begin{equation}
{\delta T_e \over T_e} \simeq -{2 \over 3}\Psi_e. \label{deltaT2}
\end{equation}
We may rewrite equation (\ref{deltaT1}), where without loss of
generality we put $\Psi_r=0$. This equation is valid if the
observation is made for regions with angular scales containing the
horizon ($\vartheta \gtrsim 2^\circ$) in recombination epoch.
Finally we recover the Sachs-Wolfe result
\begin{equation}
{\delta T_r \over T_r}={1 \over 3}\Psi_e + 2 \int
\limits_e^r{\partial \Psi \over \partial \eta}dw.
\end{equation}
The second term is called the {\em integrated Sachs-Wolfe effect}.
As it was expected, this expression for the Sachs-Wolfe effect is
the same as the one obtained for FLRW universes, for the same
order of approximation, and for adiabatic initial conditions.

\section{Study of the behavior of the ${\cal W}$, ${\dot{\cal W}}$, $\Sigma$ and $\dot \Sigma$}

As mentioned earlier in this article, we have two
scalar, ${\cal W}$ and $\Sigma$ or its square, we can
say how 'close' to FLRW are our models. When not
only one, but two parameters are close to zero, can
say that our models are `close' to FLRW
\cite {Nilsson}. Thus, we study the behavior of both, and
as their time derivatives. As for our models, with
perfect fluid, the magnetic part of Weyl is null we obtain for
${\cal W}$ from (\ref{Weyl}), after some simplification
\begin{equation}
{\cal W}=\frac{1}{3H^2}\left( \frac{\ddot a}{a}- \frac{\ddot b}{b}
\right), \label{W}
\end{equation}
or also
\begin{equation}
{\cal W}= \frac{1}{6}\frac{H_{b_0}^2}{H^2}\left[
\frac{\Omega_{M_0}}{2} \left( \frac{3}{y^3}- \frac{1}{xy^2}
\right)+ \Omega_{\Lambda_0} \frac{1}{xy^2}-
\frac{H_{a_0}}{H_{b_0}} \frac{1}{xy^2} \right], \label{W4}
\end{equation}
where $H$ is, in our case, the average value of two
Hubble parameter, $H=(\dot a/a +2 \dot b/b)/3$. Thus, we can
deriving both terms in the equation above, obtaining
\begin{figure}[h]
\centerline{\epsfig{file=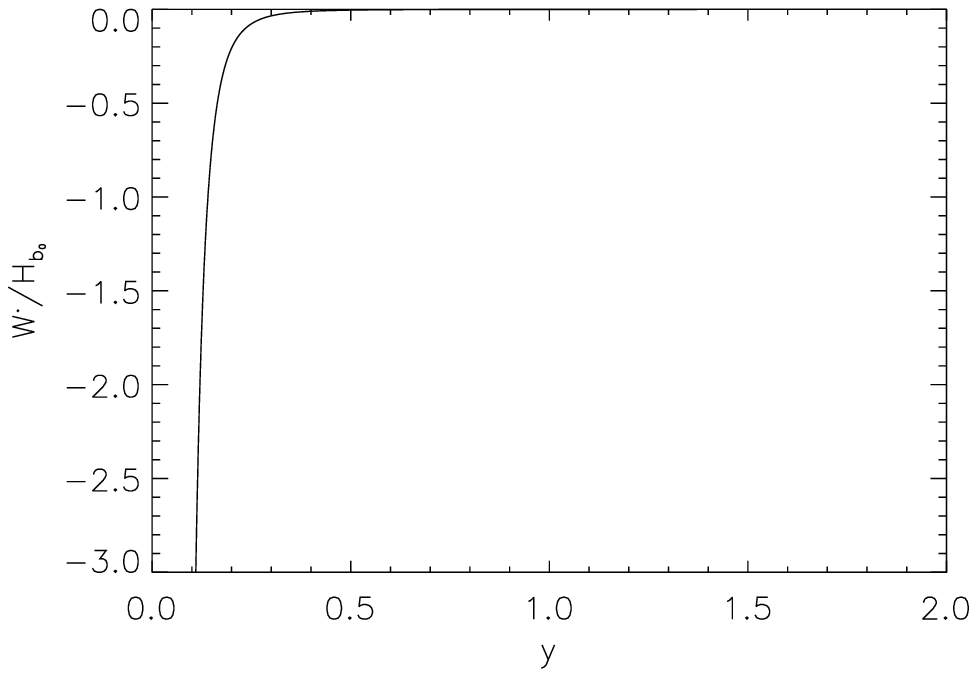,width=6.cm, angle=0}
\epsfig{file=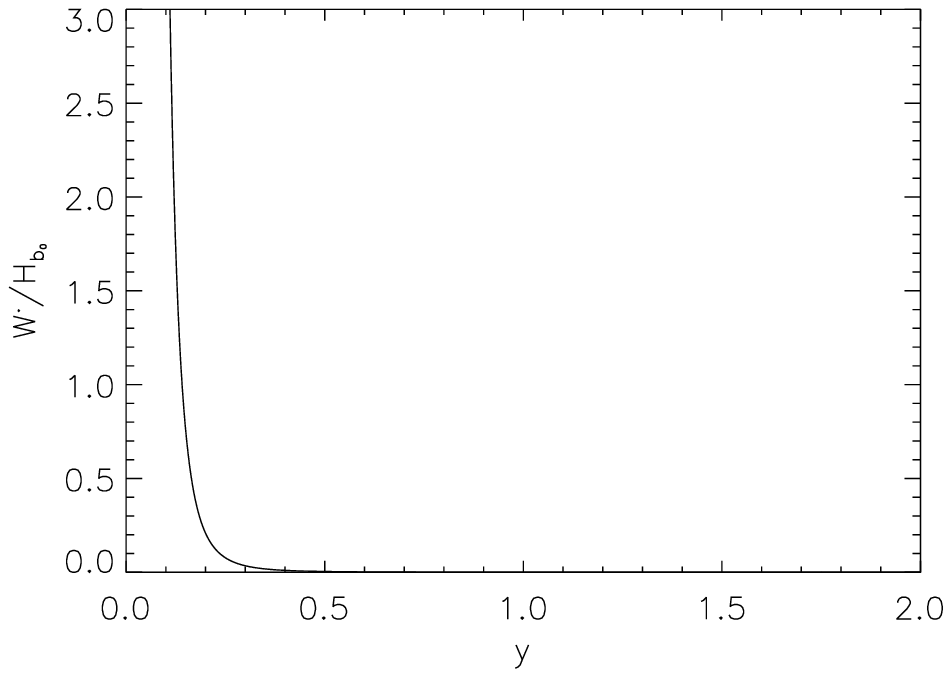,width=6.cm, angle=0}}
\caption{\footnotesize On the left: Variation of $\dot{\cal W}$ relative to scale factor
$y=b(t)/b_0$ in Kantowski-Sachs model, with the values
$\Omega_{M_0}=0.3$ and $\Omega_{\Lambda_0}=0.7 +1 \times 10^{-3}$.
It is is clear the asymptotic behavior of the $\dot{\cal W}$
($\dot{\cal W} \rightarrow -\infty$), when $y \rightarrow 0$,
that is, when we go back in time. Unlike $\dot {\cal W}
\rightarrow 0$ for $y>1$ values. Note that $y=1$
corresponds to the current instant of the Universe.
\label{fig1}}
\caption{\footnotesize On the right: Variation of $\dot {\cal W}$ relative to scale factor
$y=b(t)/b_0$ in Bianchi type-III model, with the values
$\Omega_{M_0}=0.3$ and $\Omega_{\Lambda_0}=0.7 -1 \times 10^{-3}$.
We see from the figure that $\dot{\cal W} \rightarrow +\infty$ when
$t \rightarrow 0$. For $y>1$ values $\dot {\cal W}$ converges
quickly to zero. Apart from an opposite sign, $\dot {\cal W}$
has a very similar behavior in both models.
\label{fig2}}
\end{figure}

\begin{eqnarray}
\frac{\dot {\cal W}}{H_{b_0}} &=& -\frac{\left( \frac{\ddot x}{x}+
2\frac{\ddot y}{y} -\frac{\dot x^2}{x^2}- 2\frac{\dot y^2}{y^2}
\right)}{\left( \frac{\dot x}{x}+ 2\frac{\dot y}{y} \right)} {\cal
W} + \left[ \frac{\Omega_{M_0}}{2}\left( -\frac{9}{y^3}\frac{\dot
y}{y} + \frac{1}{xy^2}\frac{\dot x}{x}+\frac{2}{xy^2}\frac{\dot
y}{y}
\right) +  \right. \nonumber \\
& & \left. \Omega_{\Lambda_0} \left( -\frac{1}{xy^2}\frac{\dot
x}{x} -\frac{2}{xy^2}\frac{\dot y}{y} \right)+
\frac{H_{a_0}}{H_{b_0}} \left( \frac{1}{xy^2}\frac{\dot x}{x}
\frac{2}{xy^2}\frac{\dot y}{y} \right) \right],
\end{eqnarray}
\begin{figure}[h]
\centerline{\epsfig{file=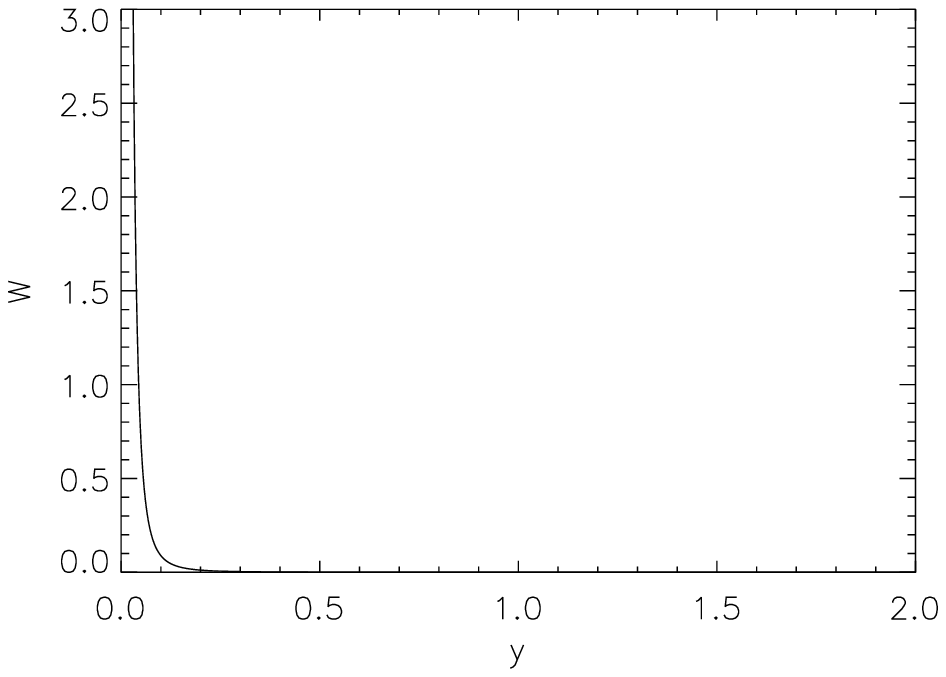,width=6.cm, angle=0}
\epsfig{file=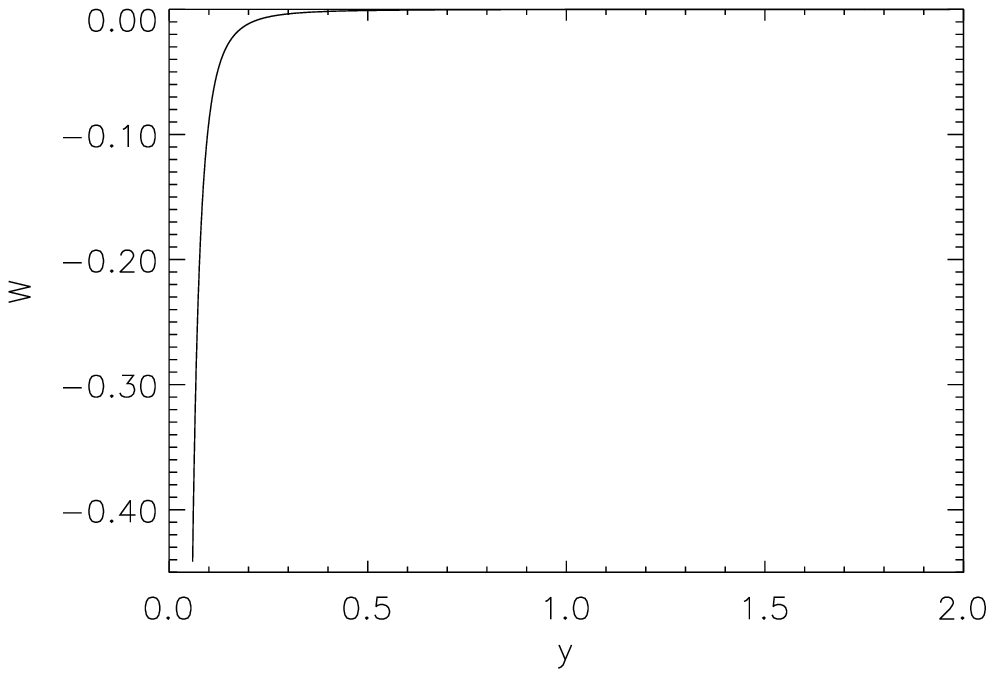,width=6.cm, angle=0}}
\caption{\footnotesize On the left: Variation of ${\cal W}$ against the scale factor
$y=b(t)/b_0$ in Kantowski-Sachs model, with the values $\Omega_{M_0}=0.3$
and $\Omega_{\Lambda_0}=0.7+1 \times 10^{-3}$. It is seen in this figure
that ${\cal W}$ remains very close to zero for a certain
period, but from a given value of $y$ close to zero, this
grows asymptotically. The $y$ value from which their growth
is asymptotic will be more close to zero as closer to the unit
is the sum of the two density parameters.
For $y>1$ values ${\cal W}$ vanishes rapidly.
\label{fig3}}
\caption{\footnotesize On the right: Variation of ${\cal W}$ against the scale factor
$y=b(t)/b_0$ in Bianchi type-III model, with the values $\Omega_{M_0}=0.3$ and
$\Omega_{\Lambda_0}=0.7 -1 \times 10^{-3}$. We see in the figure that
${\cal W}$ remains very close to zero for a certain
period, but from a given value and $y$ close to zero, $y$
diverges asymptotically. The $y$ value from which their
growth is asymptotic will be more close to zero as
closer to the unit is the sum of the two density parameters.
The ${\cal W}$ value in modulus has a very similar behavior
in both models. For $y>1$ values, ${\cal W}$ rapidly vanishes.
\label{fig4}}
\end{figure}

Using the equations
\begin{equation}
\dot y=\pm H_{b_{0}}\sqrt{\Omega_{M_{0}} \left(\frac{1}{y}-1
\right)+ \Omega_{\Lambda_{0}}(y^2-1)+1}, \label{eqy}
\end{equation}
and
\begin{equation}
\dot
x=H_{b_{0}}~\frac{\frac{\Omega_{M_{0}}}{2}\left(1-\frac{x}{y}\right)
+\Omega_{\Lambda_{0}}\left(-1+xy^2\right)+\frac{H_{a_{0}}}{H_{b_{0}}}}
{y\sqrt{\Omega_{M_{0}}\left(\frac{1}{y}-1\right)+\Omega_{\Lambda_{0}}
(y^2-1)+1}},
\label{eqx}
\end{equation}
respectively (2.12) and (2.22) obtained in Aguiar \& Crawford
\cite{Aguiar} we integrate
numerically in time the equations ${\dot{\cal W}}$ and
${\cal W}$ and we draw some graphs. For the Kantowsky-Sachs model
we chose the values $\Omega_ {M_0}=0.3$ and $\Omega_{\Lambda_0}=0.7 +1
\times 10^{-3}$ and in Bianchi type-III model we chose
$\Omega_ {M_0}=0.3$ and $\Omega_{\Lambda_0 }=0.7 -1 \times 10^{-3}$.
It is clearly seen in the figures that if we walk backwards in time
we see that $\dot{\cal W}$ diverges to $-\infty$ in the
Kantowsky-Sachs model and $+\infty$ in Bianchi type-III model. This
behavior reinforces the fact that ${\cal W}$ diverge when we go back
in time. In the Kantowski-Sachs model ${\cal W} \rightarrow +\infty$
when $t \rightarrow 0$ and in Bianchi type-III model ${\cal W}
\rightarrow -\infty$ when $t \rightarrow 0$. From equation (\ref{W4}),
and again using the equations (2.12) and (2.22) in Aguiar \& Crawford
\cite{Aguiar}, we see that ${\cal W}$ goes to zero when $x$ and
$y$ goes to infinity. If we want to relate $\dot{\cal W}$
with ${\cal W}$, we see once again that in both these models the
two scalars diverge when we walk back in time, although
$\dot{\cal W}$ diverges more sharply than ${\cal W}$.
\begin{figure}[h]
\centerline{\epsfig{file=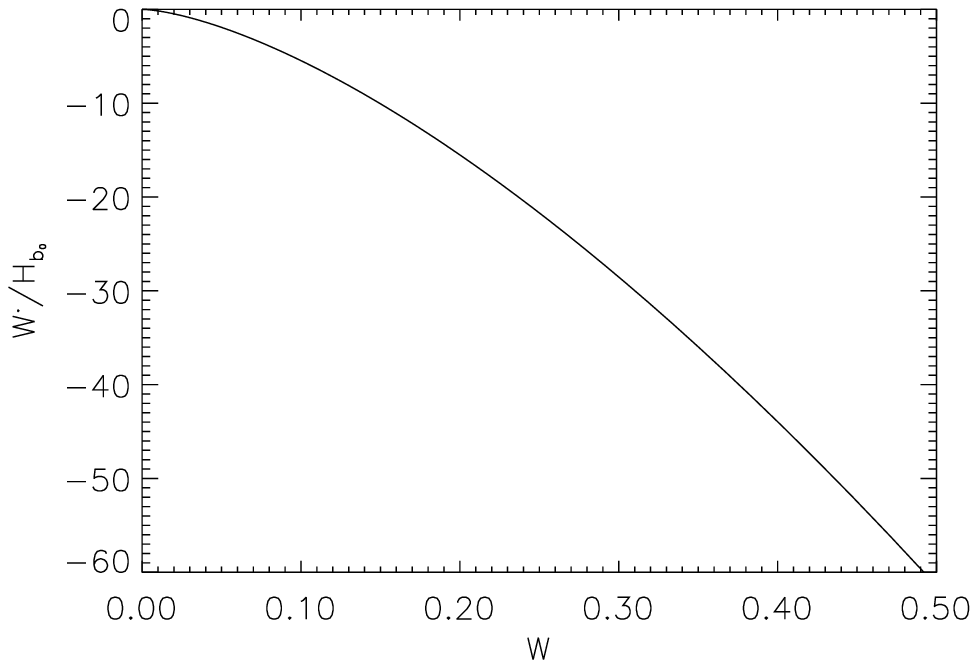,width=6.cm, angle=0}
\epsfig{file=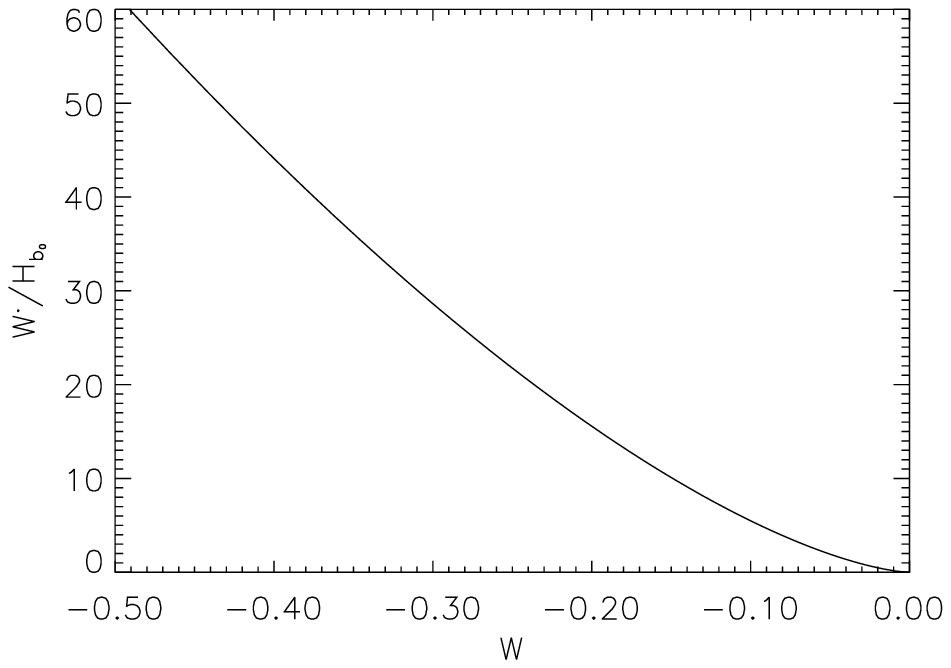,width=6.cm, angle=0}}
\caption{\footnotesize On the left: Variation of $\dot {\cal W}$ depending on ${\cal W}$ to
Kantowski-Sachs model, with values $\Omega_{M_0}=0.3$ and
$\Omega_{\Lambda_0}=0.7 +1 \times 10^{-3}$. As we go back in time
both scalar diverge, and $\dot{\cal W}$ a more pronounced divergence.
When $t \rightarrow \infty$, $\dot {\cal W}$ and ${\cal W}$ vanishes.
\label{fig5}}
\caption{\footnotesize On the right: Variation of $\dot {\cal W}$ depending on ${\cal W}$ to
Bianchi type-III model, with values $\Omega_{M_0}=0.3$ and
$\Omega_{\Lambda_0}=0.7-1 \times 10^{-3}$. As we go back in time
both scalar values diverge although by symmetrical values
to the Kantowski-Sachs model. When $t \rightarrow \infty$,
$\dot {\cal W}$ and ${\cal W}$ vanishes.
\label{fig6}}
\end{figure}

Let us study then the behavior of $\Sigma$. From the
equation (\ref {Si}) the expression of $\Sigma$ for our models,
with perfect fluid is
\begin{equation}
\Sigma = \frac{1}{3}\frac{1}{H} \left( \frac{\dot a}{a}
-\frac{\dot b}{b} \right).
\end{equation}
Deriving both members we obtain easily
\begin{equation}
\dot \Sigma = -\frac{\left( \frac{\ddot x}{x}+ 2\frac{\ddot y}{y}
-\frac{\dot x^2}{x^2}- 2\frac{\dot y^2}{y^2} \right)}{\left(
\frac{\dot x}{x}+ 2\frac{\dot y}{y} \right)} \Sigma +
\frac{\frac{\ddot x}{x}-\frac{\ddot y}{y}- \frac{\dot x^2}{x^2}
+\frac{\dot y^2}{y^2}}{\left( \frac{\dot x}{x}+ 2\frac{\dot y}{y}
\right)}.
\end{equation}
Using back to equations (2.12), (2.22) from Aguiar \& Crawford
\cite {Aguiar} we have integrated
numerically in order to time the equations $\dot \Sigma$ and
$\Sigma$ and got some graphs illustrative of these behavior
Also the scalar $\Sigma$ has a behavior highly divergent
when we stepped back in time and highly convergent to
zero when we consider values of scale factors $y \sim 1$ in
both models. This behavior is strongly supported
by the behavior of its derivative $\dot \Sigma$.

Both in the study of the scalar ${\cal W}$ as the study of $\Sigma$,
we use density parameter values $| \Omega_{M_0}+\Omega_{\Lambda_0}-1|
=1 \times 10^{-3}$, just to illustrate the qualitative behavior of both models.
However, if we choose values for density parameters of the order of
$| \Omega_{M_0}+\Omega_{\Lambda_0}-1| \simeq 10^{-9}$ and making $H_{a_0}/H_ {b_0} \pm 3.6
\times 10^{-9}$ we obtain the following values:
\begin{equation}
{\cal W}_0 \sim 2 \times 10^{-10} ~~~~~~\mbox{e}~~~~~~ {\cal
W}_{ls} \sim 3.8 \times 10^{-1}
\end{equation}
and
\begin{equation}
\Sigma_0 \sim -3.6 \times 10^{-9} ~~~~~~\mbox{e}~~~~~~ \Sigma_{ls}
\sim -4.4 \times 10^{-5}
\end{equation}
in Kantowski-Sachs model and
\begin{equation}
{\cal W}_0 \sim -2 \times 10^{-10} ~~~~~~\mbox{e}~~~~~~ {\cal
W}_{ls} \sim -5.6 \times 10^{-1}.
\end{equation}
and
\begin{equation}
\Sigma_0 \sim 3.6 \times 10^{-9} ~~~~~~\mbox{e}~~~~~~ \Sigma_{ls}
\sim 6.5 \times 10^{-5}
\end{equation}
in Bianchi type-III model.

The behavior of these scalar clearly shows that we can
keep close to our models from FLRW behavior
provided that the appropriate density parameter settings
and Hubble parameters are close to the present time.
However, going back to ``last scattering" time these
models exhibit strongly anisotropic behavior.
\begin{figure}[h]
\centerline{\epsfig{file=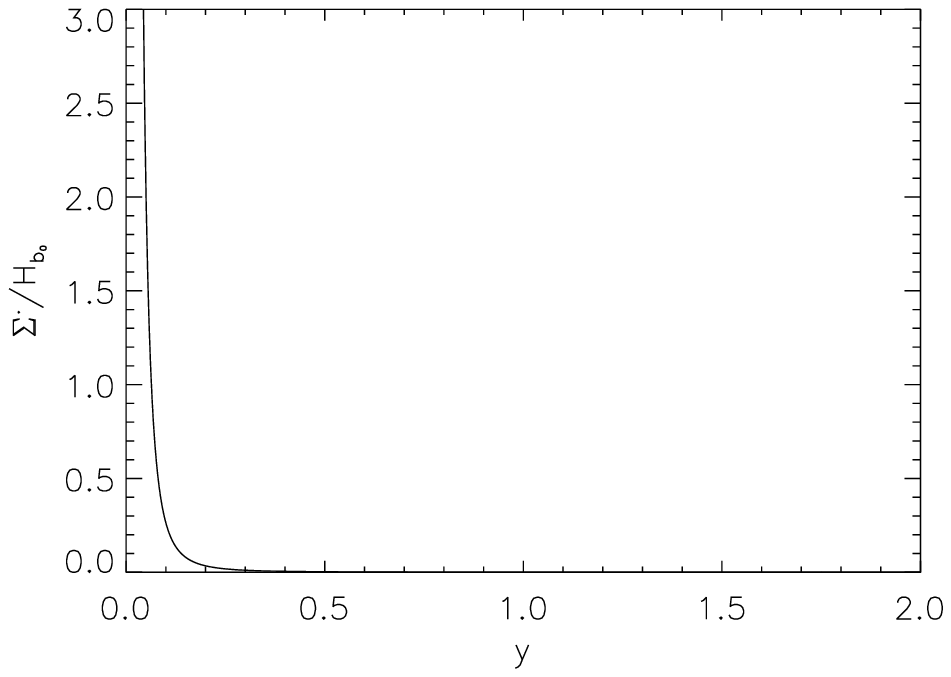,width=6.cm, angle=0}
\epsfig{file=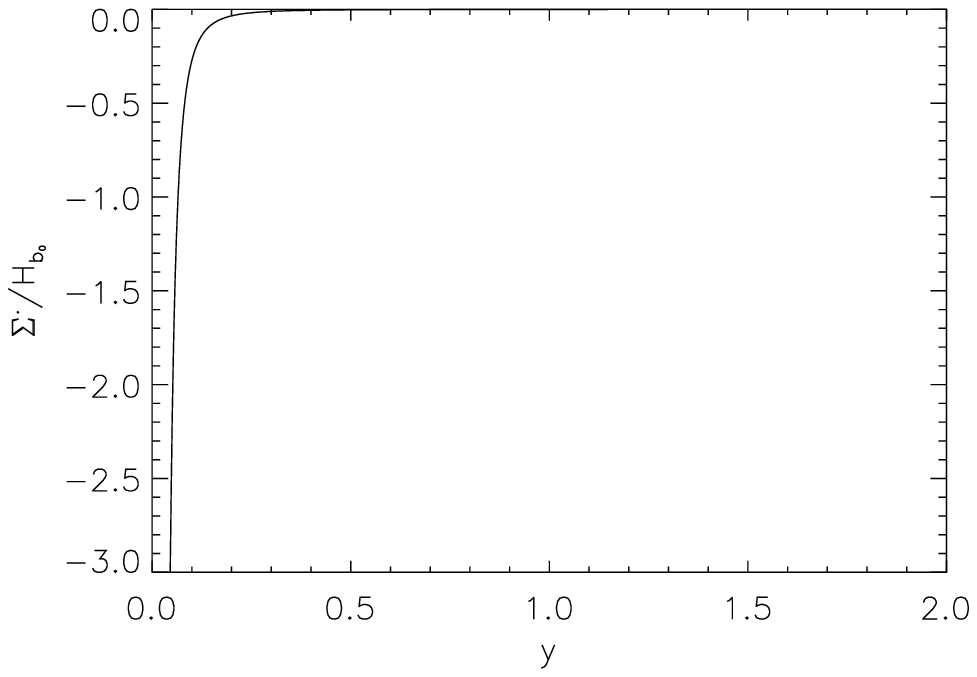,width=6.cm, angle=0}}
\caption{\footnotesize On the left: Variation of $\dot \Sigma$ against the scale factor
$y=b(t)/b_0$ in Kantowski-Sachs model, with values $\Omega_{M_0}=0.3$
and $\Omega_{\Lambda_0}=0.7+1 \times 10^{-3}$. It is clear the
asymptotic behavior of $\dot \Sigma$ ($\dot \Sigma \rightarrow -\infty$),
when $y \rightarrow 0$, ie when we go back in time. Unlike
$\dot {\cal W} \rightarrow 0$ for $y>1$ values. Note that $y=1$ corresponds
to the present time in Universe.
\label{fig7}}
\caption{\footnotesize On the right: Variation of $\dot \Sigma$ against the scale factor
$y=b(t)/b_0$ in Bianchi type-III model, the values $\Omega_{M_0}=0.3$ and
$\Omega_{\Lambda_0}=0.7 -1 \times 10^{-3}$. We see from the figure that
$\dot \Sigma  \rightarrow +\infty$ when $t \rightarrow 0$. For $y>1$
values $\dot \Sigma$ converges quickly to zero. We see that $| \dot \Sigma|$
has a behavior very similar in both models.
\label{fig8}}
\end{figure}
\begin{figure}[h]
\centerline{\epsfig{file=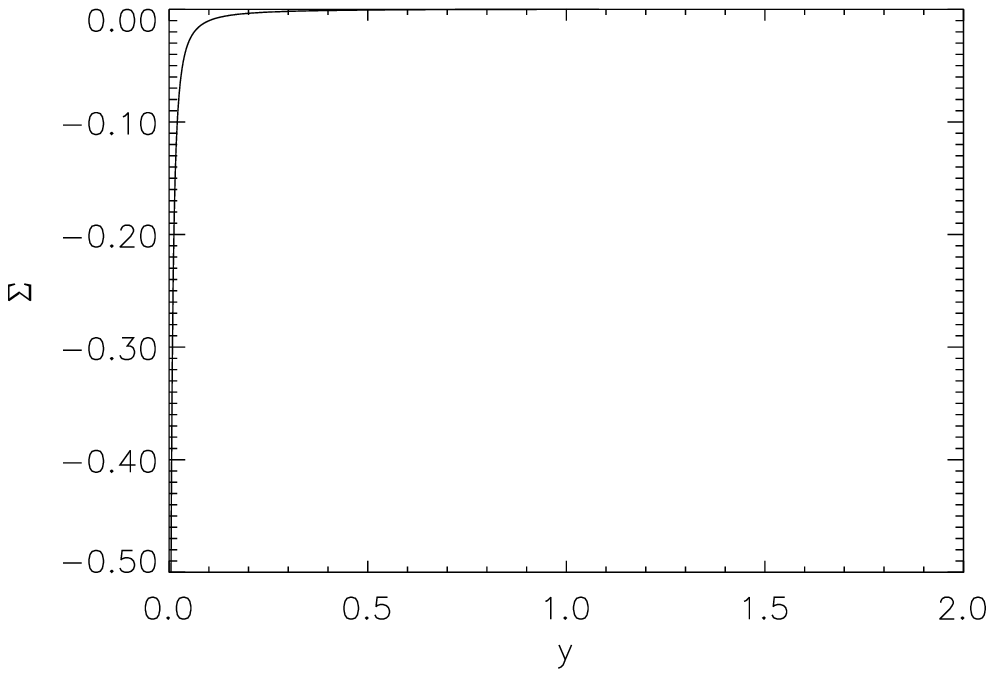,width=6.cm, angle=0}
\epsfig{file=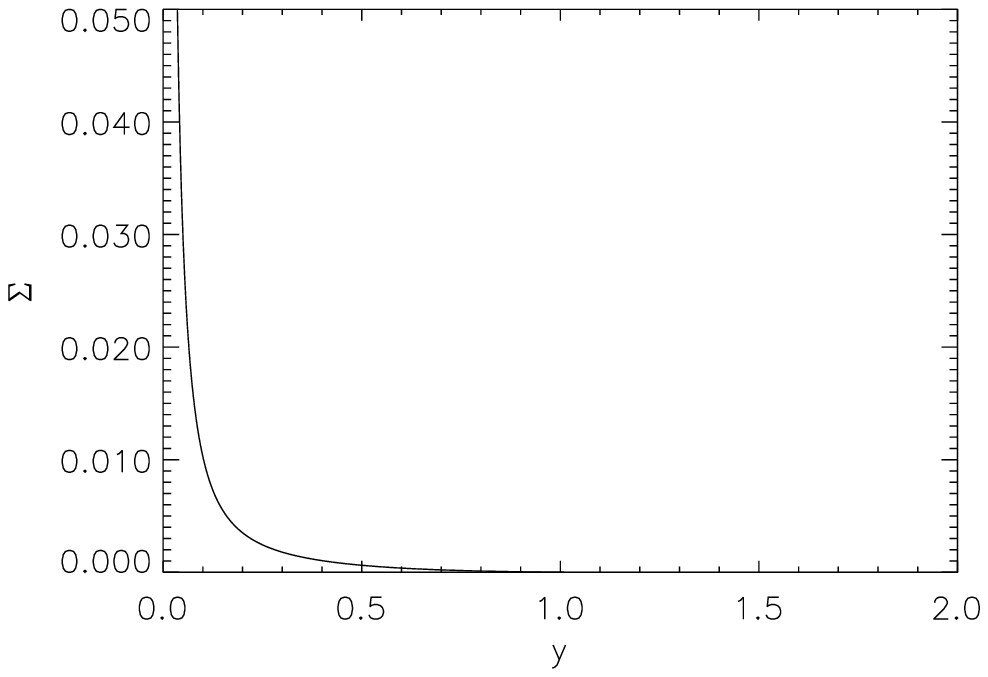,width=6.cm, angle=0}}
\caption{\footnotesize On the left: Variation of $\Sigma$ against the scale factor
$y=b(t)/b_0$ in Kantowski-Sachs model, with values $\Omega_{M_0}=0.3$
and $\Omega_{\Lambda_0}=0.7 +1 \times 10^{-3}$. It is seen in this figure
$\Sigma$ remains very close to zero for a certain period,
but from a given value of $y$ close to zero, this increases
asymptotically. The value $y$ from which their growth is asymptotic will
be more close to zero as closer to the unit is the sum of the two density parameters.
For $y>1$ values $\Sigma$ vanishes quickly.
\label{fig9}}
\caption{\footnotesize On the right: Variation of $\Sigma$ against the scale factor
$y=b (t)/b_0$ in Bianchi type-III model, the values $\Omega_{M_0}=$ and
$\Omega_{\Lambda_0}=0.7 -1 \times 10^{-3}$. It is seen in this figure
$\Sigma$ remains very close to zero for a certain period,
but from a given value of $y$ close to zero, this diverges
asymptotically. The $y$ value from which their growth
is asymptotic will be more close to zero as
closer to the unit is the sum of the two density parameters.
The $| \Sigma|$ has a behavior very similar in both models.
For $y>1$ values $\Sigma$ quickly vanishes.
\label{fig10}}
\end{figure}
\begin{figure}[h]
\centerline{\epsfig{file=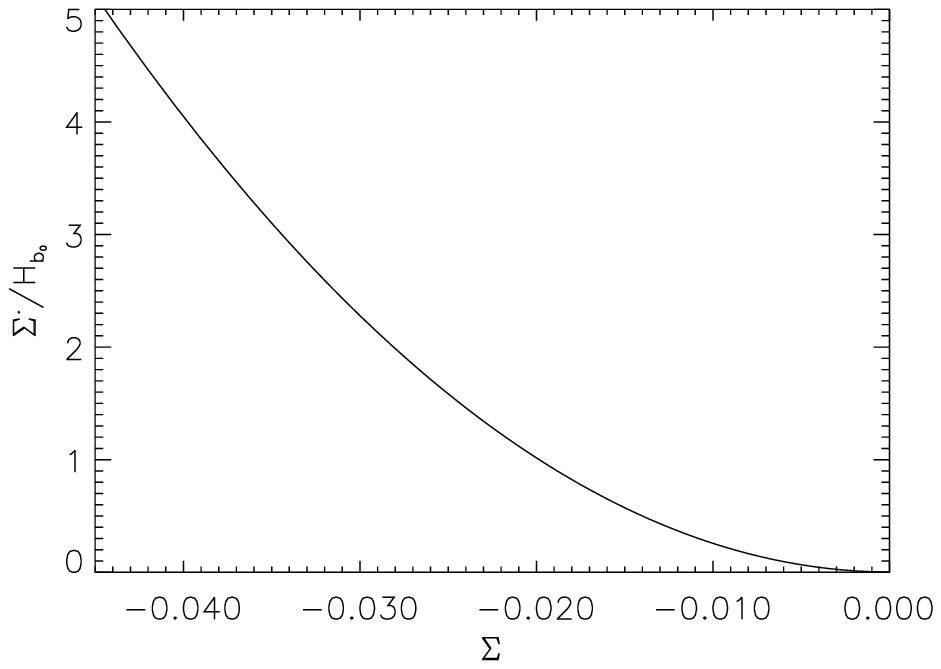,width=6.cm, angle=0}
\epsfig{file=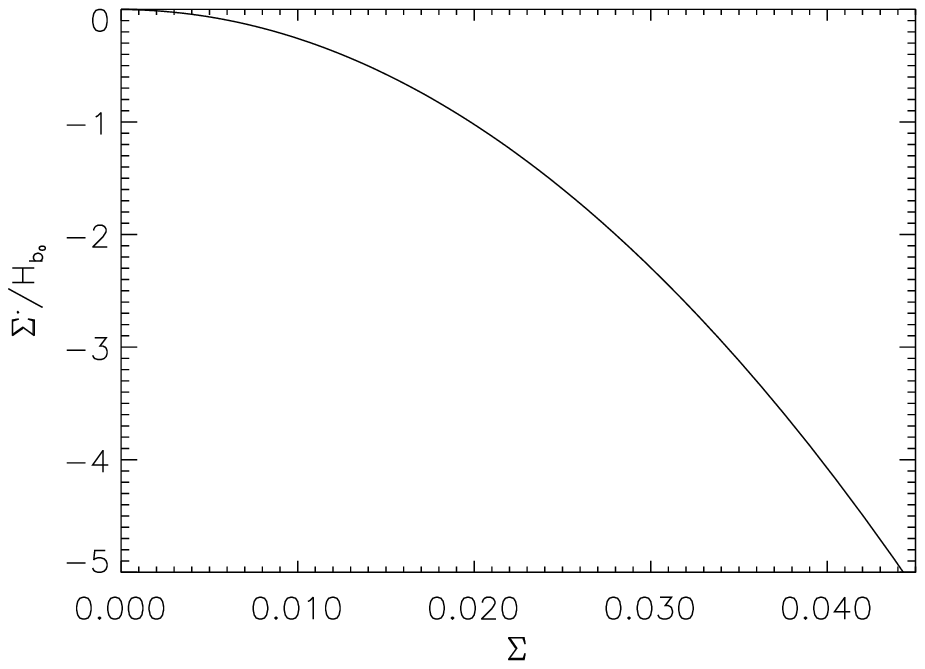,width=6.cm, angle=0}}
\caption{\footnotesize On the left: Variation of $\dot \Sigma$ as a function of $\Sigma$, to
Kantowski-Sachs model, with values $\Omega_{M_0}=0.3$ and
$\Omega_{\Lambda_0}=0.7+1 \times 10^{-3}$.
When we go back in time both scalar diverge, and $\dot \Sigma$
as a more pronounced divergence. When $t \rightarrow \infty$,
$\dot \Sigma$ and $\Sigma$ tend to zero.
\label{fig11}}
\caption{\footnotesize On the right: Variation of $\dot \Sigma$ as a function of $\Sigma$, to
Bianchi type-III model, with values $\Omega_{M_0}=0.3$ and
$\Omega_{\Lambda_0}=0.7-1 \times 10^{-3}$.
When we go back in time both scalar diverge although by symmetrical
values of Kantowski-Sachs model. When $t\rightarrow \infty$,
$\dot \Sigma$ and $\Sigma$ tend to zero.
\label{fig12}}
\end{figure}

\section{Concluding Remarks}

We stress once more we should bear in mind that the assumption
$H_a \simeq H_b$ does not imply, by itself, an isotropic or even
an almost isotropic metric, as it is expressed by the growth of
Weyl term in Equations (\ref{Weyl}) and (\ref{Si}), when we go
back in time to the last scattering epoch. Although the $\Sigma$
term remains at a low value from the present ($\Sigma_0 \sim -3.6
\times 10^{-9}$) to the last scattering epoch ($\Sigma_{ls} \sim
-4.4 \times 10^{-5}$), the Weyl term,
grows from ${\cal W}_0 \sim 2 \times 10^{-10}$ at the present
to ${\cal W}_{ls} \sim 3.8 \times 10^{-1}$ at the last scattering
time. This shows the anisotropic character of these models in the
past. Summarizing, $\Sigma$ grows about $1.2 \times 10^{4}$, while
${\cal W}$ grows about $1.9 \times 10^{9}$, when we go back in
time ($t_0 \rightarrow t_{ls}$). Even though we impose a high
level of isotropy at present time, the anisotropic behavior of
these models comes forward as we go back in time. Nevertheless,
the growth of ${\cal W}$ term does not affect decisively the first
order computation of $\delta T_r/T_r$ term.

These results could lead us to conclude that the accuracy in
density parameter settings and Hubble parameters ($H_a$ e $H_b$)
introduced was so high that these models were
`transformed' in isotropic models and therefore of no interest
study. The behavior of the scalar ${\cal W}$ and
$\Sigma$ and their time derivatives has allowed to
conclude that in fact these models can display a high level
isotropy for a period of time too high since
that $ {\cal W}$ and $\Sigma$ can remain approximately zero.
But if we go back in time to very early times, these
scalar tends to infinity, regardless of the precision
to choose the parameters of Hubble and density, which shows the
anisotropic nature of these models.

Since the obtained expression (for Sachs-Wolfe effect) is the same
as the one given for FLRW flat model, we may conclude that, these
anisotropic models are also good candidates to the description of
observed Universe provided we may assume $H_{a_0} \simeq H_{b_0}$
(taking into account the upper bound on the present value of the
shear parameter imposed by COBE observations) and a particular
choice of the density parameters: $\Omega_0 + \Omega_{\Lambda_0}
\simeq 1$, (see \cite{Aguiar2}). This is another step taken in the
same direction as in \cite{Aguiar}. This is also in agreement with
another previous result: it is not possible to distinguish a
Kantowski-Sachs model from the FLRW models, with the classical
tests of Cosmology, if the Hubble parameters along the orthogonal
directions are assumed to be approximately equal \cite{Henriques}.

In future work we intend to use Plank satellite data \cite{Planck}
and interferometer AMIBA \cite{Amiba}, which will provide much
better resolution and which will require to consider the integrated
Sachs-Wolfe effect.

In conclusion, the observation of Sachs-Wolfe effect plateau does not
permit us to distinguish between FLRW models and the anisotropic
Kantowski-Sachs and Bianchi type-III models. To investigate this
in more detail, it is necessary to consider and process the data
from Plank \cite{Planck} and AMIBA \cite{Amiba} projects to
regions smaller than the horizon at the last scattering
($\ell >100$, $\vartheta <1^\circ$). Within this region of
multipoles, perturbations are model dependent. Only with this
information we may conclude finally whether our Universe
can be modeled by one of these anisotropic models. This will be
the purpose of further work.

\section*{Acknowledgements}

The authors thank A. Barbosa Henriques, Ant\'onio da Silva and
Andrew Liddle for useful discussions and comments. This work was
supported in part by grants BD 971 and BD/11454/97 PRAXIS XXI,
from JNICT, CERN/P/FIS/40131/2000 Project
and by PEst-OE/FIS/UI2751/2011 Pro\-ject from FCT.

\end{document}